\documentclass{pic2012}
\usepackage{mdwlist}
\usepackage{subfigure}
\def\runauthor{Arun Kumar, Kirti Ranjan on behalf of the CMS collaboration}
\def\shorttitle{$H\rightarrow ZZ\rightarrow 2l2\nu$ Analysis in CMS}
\begin{document}

\title{Search for the standard model Higgs boson in the $H\rightarrow ZZ\rightarrow 2l2\nu$ decay channel in the $\it{CMS}$ Experiment.}

\author{Arun Kumar, Kirti Ranjan \\   on behalf of the $\it {CMS}$ $\it{collaboration}$}

\address{University of Delhi\\
Delhi-110007, India\\
E-mail: arun.kumar@cern.ch \\
kirti.ranjan@cern.ch }

\maketitle

\abstracts{ A search for the Standard Model Higgs boson in the $H\rightarrow ZZ\rightarrow 2l2\nu$ decay channel, where l = e or $\mu$, in pp collisions at a center-of-mass energy of both 7 and 8 TeV is presented. The data were collected at the LHC, with the CMS detector, and correspond to an integrated luminosity of 5.0 $fb^{-1}$ at 7 TeV and 5.0 $fb^{-1}$ at 8 TeV. The search is optimized separately for the vector boson fusion and the gluon fusion production processes.No significant excess is observed above the background expectation, and upper limits are set on the Higgs boson production cross section. The presence of the standard model Higgs boson with a mass in the $278-600 GeV/c^{2}$ range is excluded at 95\% confidence level.}

\section{Introduction} The Standard model (SM) of particle physics accommodates essentially the majority of high-energy experimental data. Two major remaining questions are related with the origin of mass for fundamental particles and the behavior of the $W_{L}W_{L}$ scattering cross section at high energies. Within the SM, mass of the fermions and bosons arise from the spontaneous breaking of electroweak symmetry. The existence of the associated field quantum, the Higgs boson, is yet to be experimantally established. It is also expected to regularize the behavior of the $W_{L}W_{L}$ cross section. The search for the SM Higgs boson and its discovery or exclusion are central to the goals of the experiments at the LHC. The primary production mechanism of the SM Higgs at the LHC is gluon fusion with the small, but distinct contribution from vector boson fusion. \\
An excess is found at the low end of the SM Higgs mass ($\sim$ 125 $GeV/c^{2}$) spectrum explored by both the LHC and the Tevatron experiments, so the main interest in the Higgs searches is currently focused around that mass region. This note presents a search for the SM Higgs boson in the $H\rightarrow ZZ\rightarrow 2l2\nu$ channel (where $\it{l}$ refers to either $\it{e}$ or $\mu$). This decay mode is essentially sensitive in the high-mass range above 200 $GeV/c^{2}$. Results are reported from a data sample recorded in 2011 and 2012 by the Compact Muon Solenoid (CMS) experiment \cite{CMS} at the Large Hadron Collider (LHC) corresponding to an integrated luminosity of 5.0 $fb^{-1}$ at $\surd s=$ 7 TeV and 5.0 $fb^{-1}$ at $\surd s$ = 8 TeV.

\section{Event Selection}
Since this analysis is carried out for the SM Higgs mass ($M_{H}$) above $200 GeV/c^{2}$ so $H\rightarrow ZZ\rightarrow 2l2\nu$ event is characterized by the presence of a boosted $Z$ boson decaying to $e^{+}e^{-}$ or $\mu^{+}\mu^{-}$ and large missing transverse energy ($E_{T}^{miss}$) arising from the decay of the other $Z$ boson into neutrinos. Events are selected such that there are two well-identified, isolated leptons of the same flavor ($e^{+}e^{-}$ or $\mu^{+}\mu^{-}$)  with $p_{T} > 20 GeV/c$ that have an invariant dilepton mass within a 30 $GeV/c^{2}$ window centered on the $Z$ mass.\\
Muons are detected with the silicon tracker and the muon system \cite{CMS-PAS-MUO}. Further identification criteria based on the number of hits in the tracker and muon system, the fit quality of the muon track and its consistency with the primary vertex, are imposed on the reconstructed muons. Electrons are detected in the electromagnetic calorimeter (ECAL) as energy clusters and as tracks in the tracker \cite{CMS-PAS-EGM}. These reconstructed electrons are further required to pass certain identification criteria based on the ECAL shower shape, track-ECAL cluster matching and consistency with the primary vertex. They are measured in pseudorapidity ($\eta$) range $|\eta|<$ 2.4 for muons and $|\eta|<$ 2.5 for electrons, though for electrons the transition range between the barrel and endcap, 1.4442 $< |\eta| <$ 1.566, is excluded. Additional requirements are imposed to remove electrons produced in photon conversions in the detector material. \\
Lepton isolation is defined from the sum of the flux of the momenta of the particle flow based candidates found in a cone of radius $R = \sqrt{{\Delta\eta}^2 + {\Delta\phi}^2} = 0.4$ built around each lepton, where $\phi$ is the azimuthal angle. Relative isolation sum is required to be smaller than 15\% (20\%) for electrons (muons). Isolation is corrected from pile-up effects by subtracting a product of median energy density ($\rho$)\cite{rho_cite} and effective area (derived in order to re-normalize the $\rho$ estimate to the number of pile-up interactions) from the isolation sum. For muons, this correction is achieved by subtracting half the sum of $p_{T}$ of the charged particles in the cone of interest, but with particles not originating from the primary vertex.\\
Jets are reconstructed from particle-flow candidates \cite{CMS-PAS-JME} by using the anti-$k_{T}$ clustering algorithm \cite{jhep_antikt} with cone of radius $R = 0.5$.\\
With this selection the principal backgrounds in this analysis are:
\begin{itemize*}
\setlength\listparindent{0.5in}
\setlength{\parskip}{0pt}
\setlength{\parsep}{0pt}
\setlength{\headsep}{0pt}
\setlength{\topskip}{0pt}
\setlength{\topmargin}{0pt}
\setlength{\topsep}{0pt}
\setlength{\partopsep}{0pt}
\setlength{\itemsep}{0ex}
\item{{\bf $Z$+jets}: with missing transverse energy due to jet mis-measurement and detector effects.}
\item{{\bf Non-Resonant}(i.e., events without a $Z$ resonance): top quark decays, fully leptonic $WW$ decays, $W$+jets with jet identified as lepton. } 
\item{{\bf Irreducible}: electroweak $ZZ$ pair production and fully leptonic decays of $WZ$ pairs.}
\end{itemize*}
The other selection variables which are used in this analysis are as follows-
\begin{itemize*}
\setlength\listparindent{0.5in}
\setlength{\parskip}{0pt}
\setlength{\parsep}{0pt}
\setlength{\headsep}{0pt}
\setlength{\topskip}{0pt}
\setlength{\topmargin}{0pt}
\setlength{\topsep}{0pt}
\setlength{\partopsep}{0pt}
\setlength{\itemsep}{0ex}
\item{{\bf B-tagging and soft-muon veto}: Since top quark events contain $b$-jets, this background can be suppressed by vetoing events with at least one $b$-tagged jet with transverse energy greater than 30 GeV that lies within the tracker volume ($|\eta| <$ 2.5). Top-quark decays are characterized by the presence of jets originating from $\it b$ quarks ($b$-jets), which can be tagged by using a likelihood that the tracks associated to a jet are consistent with coming from the primary vertex based on the impact parameter significance \cite{btag_cite}. In addition to the $b$-jet veto, a veto is applied on events containing ``soft muon'' with $p_{T} >$ 3 GeV/c, which is typically produced in the leptonic decay of a $\it b$ quark.}
\item{{\bf Third lepton veto}: In order to suppress the $WZ$ background in which both $W$ and $Z$ decay leptonically, events are required to have exactly two leptons with $p_{T} >$ 10 GeV/c. } 
\item{{\bf{$\Delta\phi(E_{T}^{miss},jet)$} }: To suppress background with $E_{T}^{miss}$ coming from jet mis-measurements, events are removed if the angle in the azimuthal plane between the $E_{T}^{miss}$ and the closest jet (with transverse energy $E_{T} >$ 30 GeV) is smaller than 0.5 radians.}
\item{{\bf{$\Delta\phi(l,E_{T}^{miss})$}}: In order to remove events where the lepton has been mismeasured, events are rejected if $\Delta\phi(l,E_{T}^{miss}) <$ 0.2 radian.}
\item{{\bf $E_{T}^{miss}$}: Particle flow $E_{T}^{miss}$ \cite{CMS-PAS-PFT} is used in the analysis. A stringent cut on $E_{T}^{miss}$ mainly suppresses Drell-Yan background.}
\item{{\bf Transverse mass of Higgs ($M_{T}$)}: Signal events have a narrower $M_{T}$ distribution. A two-sided cut is applied on the $M_{T}$ to further separate signal with respect to the background.}
\end{itemize*}

\section{Analysis Strategy}
This analysis has been divided into different jet multiplicity categories defined as follows:
\begin{itemize*}
\setlength\listparindent{0.5in}
\setlength{\parskip}{0pt}
\setlength{\parsep}{0pt}
\setlength{\headsep}{0pt}
\setlength{\topskip}{0pt}
\setlength{\topmargin}{0pt}
\setlength{\topsep}{0pt}
\setlength{\partopsep}{0pt}
\setlength{\itemsep}{0ex}
\item{{\bf VBF category} - In this category there is a requirement of two or more jets in forward region with $|\Delta\eta| >$ 4 between the closest jets, and a minimal invariant mass of those two jets of 500 $GeV/c^{2}$. The two leptons forming the $Z$ candidate are required to lie in between these two jets, while no other selected jets are allowed in this central region.}
\item{{\bf Gluon fusion category} - All events failing the VBF selection are divided according to the number of reconstructed jets with $p_{T} >$30 GeV/c. The two jet category is inclusive of all higher jet multiplicities.} 
\end{itemize*}
The $E_{T}^{miss}$ and $M_{T}$ selection listed in Table \ref{tab:cuts_gg} was optimized based on the expected limits. In case of VBF category a constant $E_{T}^{miss} >$70 GeV and without any $M_{T}$ requirement are used as there was no gain in sensitivity with a $M_{H}$ dependent selection.

\begin{table}[ht]
\centering
\begin{tabular}{|l|l|l|}
\hline
$M_{H} (GeV/c^{2}$) & $E_{T}^{miss}$ & $M_{T}$ \\
\hline
200 & $>$ 75 & 175-275 \\
\hline
300 & $>$ 85 & 250-375 \\
\hline
400 & $>$ 90 & 325-450 \\
\hline
500 & $>$ 90 & 400-650  \\
\hline
600 & $>$ 100 & 450-$\infty$ \\
\hline
\end{tabular}
\caption{Higgs mass-dependent selection for $E_{T}^{miss}$ and $M_{T}$ variables in the gluon fusion category.}
\label{tab:cuts_gg}       
\end{table}

\section{Background Estimation}
$ZZ/WZ$ backgrounds are modeled using Monte Carlo simulation, and are normalized to their respective NLO cross sections. The remaining backgrounds ($Z$+jets and all non-resonant ones) are estimated using control samples in data.
\subsection{{\bf $Z$+jets Estimation}}
\begin{figure}[ht]
\begin{center}
\subfigure[]{\label{fig:MET_plots-a}\includegraphics[angle=0,totalheight=.15\textheight,width=.30\textwidth]{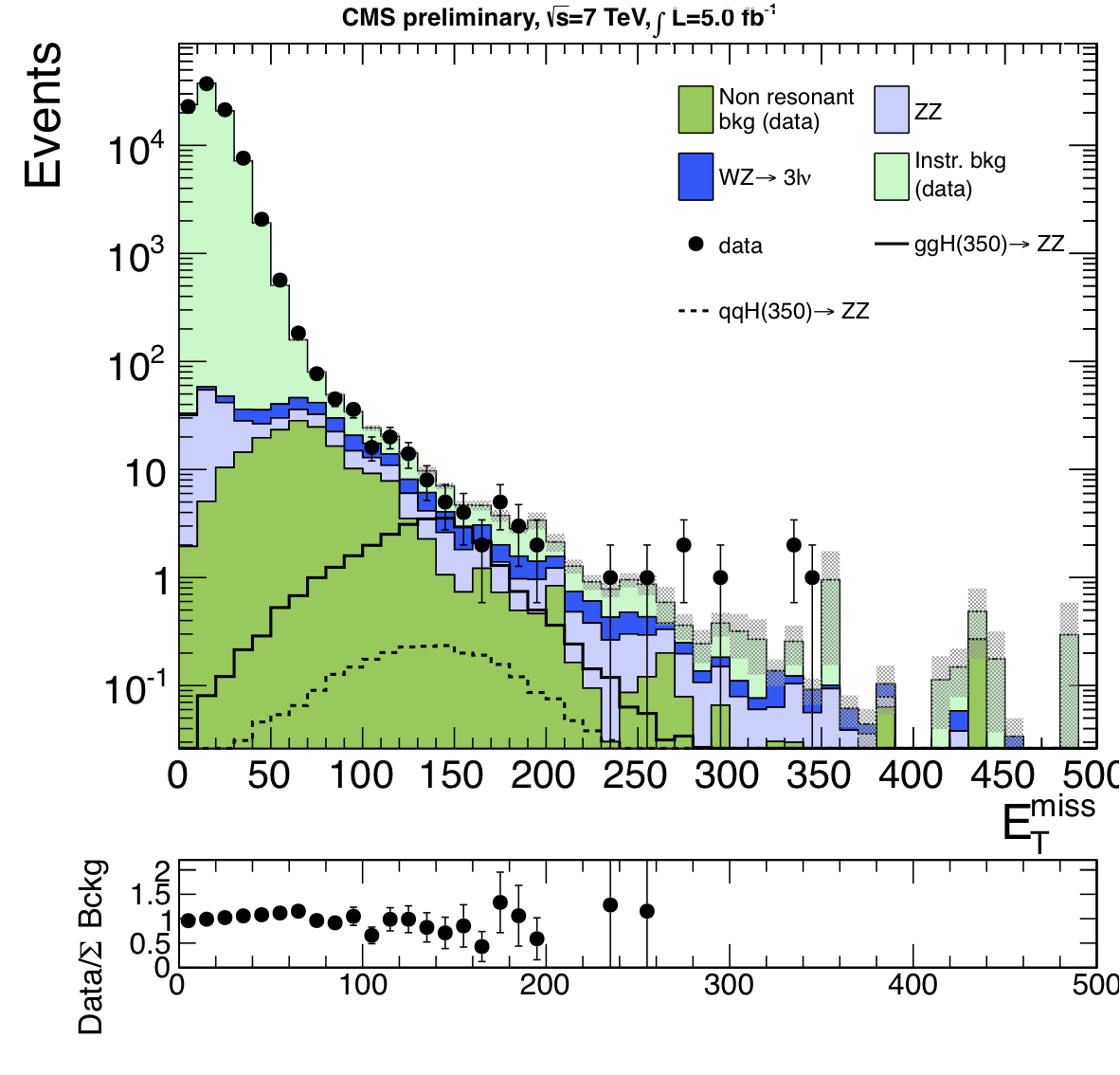}}
\subfigure[]{\label{fig:MET_plots-b}\includegraphics[angle=0,totalheight=.15\textheight,width=.30\textwidth]{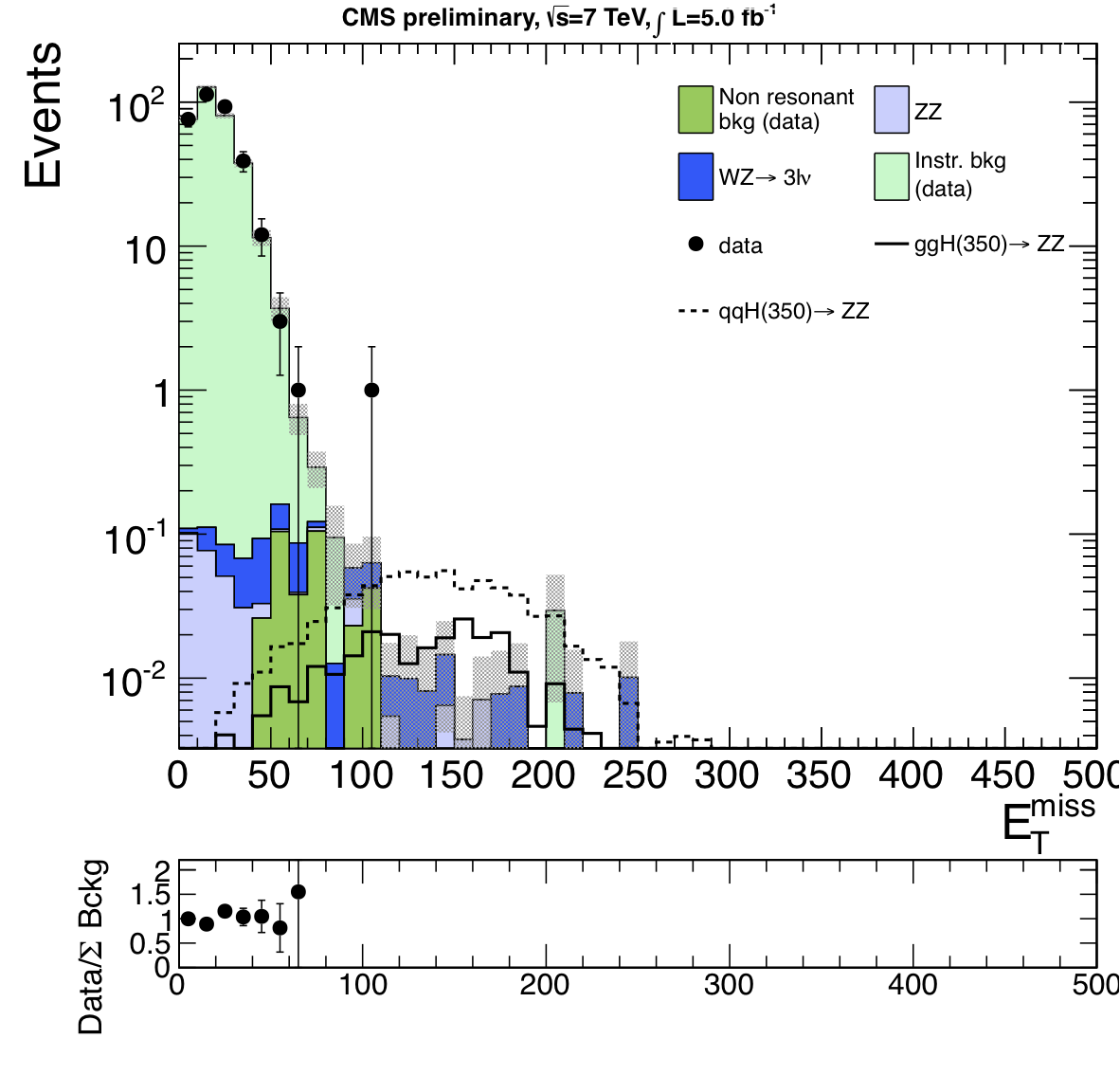}}\\
\subfigure[]{\label{fig:MET_plots-c}\includegraphics[angle=0,totalheight=.15\textheight,width=.30\textwidth]{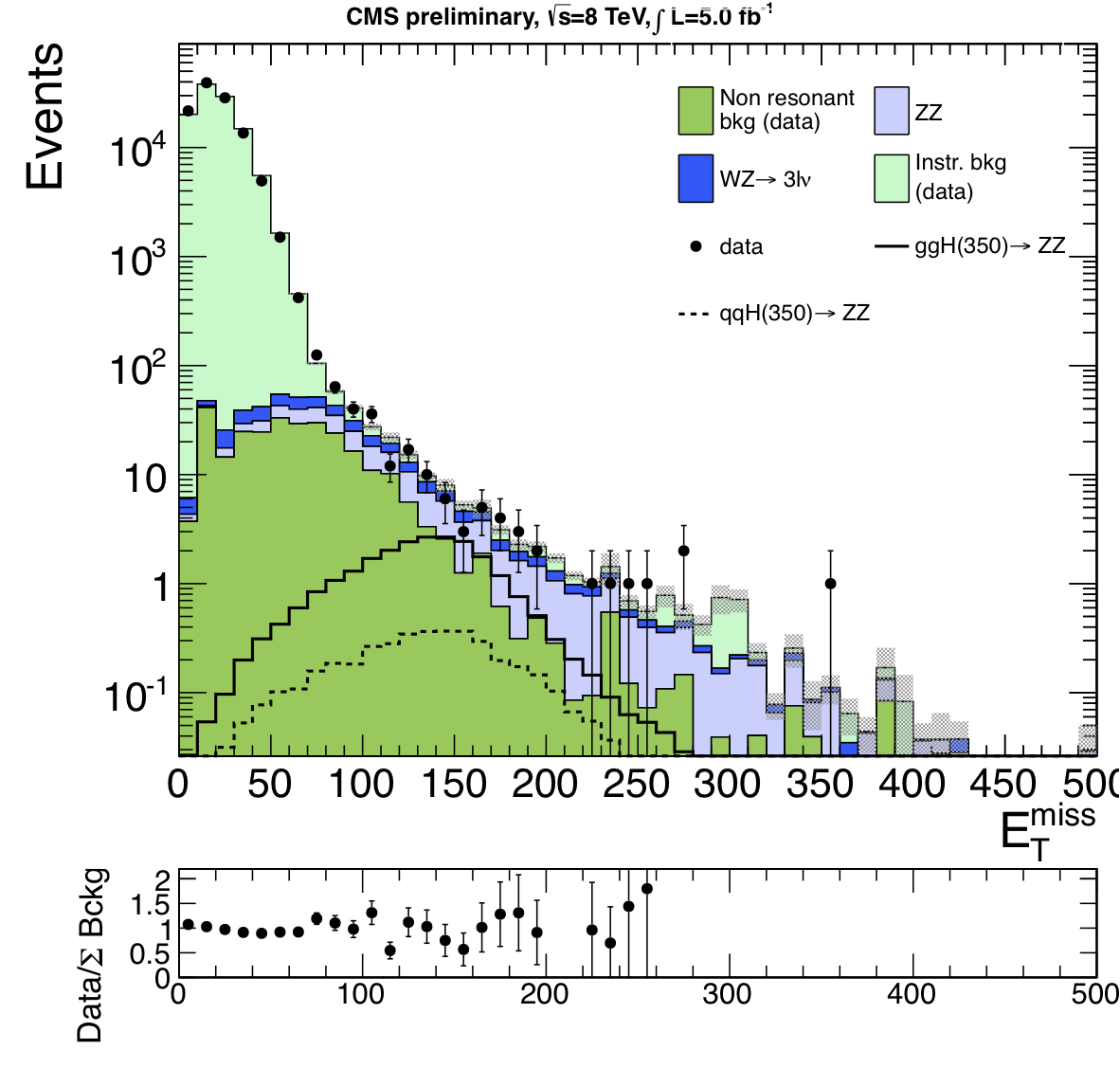}} 
\subfigure[]{\label{fig:MET_plots-d}\includegraphics[angle=0,totalheight=.15\textheight,width=.30\textwidth]{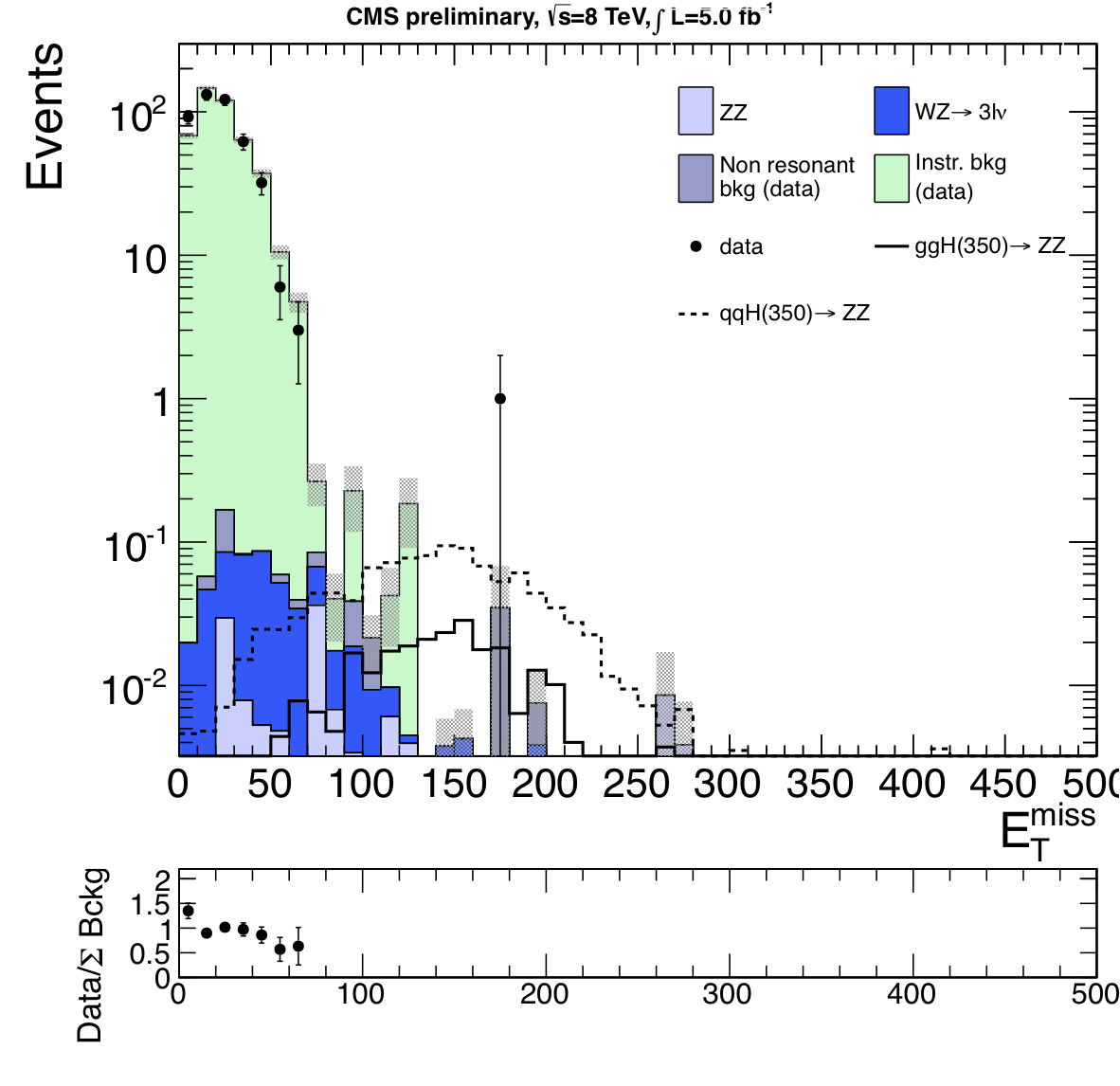}} \\
\caption{The $E_{T}^{miss}$ distribution in data compared to the estimated background from data or simulation in the gluon fusion and VBF categories. The top plots show the results for $\surd s$ = 7 TeV analysis, while the bottom plots show the same for $\surd s$ = 8 TeV. The dielectron and dimuon channels are combined. Contributins from $ZZ,WZ$,non-resonant background and $Z+jets$ background are stacked on top of each other. The $E_{T}^{miss}$ distribution in signal events for $m_{H} =$ 350 $GeV/c^{2}$ is also shown. The smaller plots show ratio of observed data to expected background events. Y-axis is zoomed in around 1 and a few outliers appear as empty bins.}
\label{fig:MET_plots}
\end{center}
\end{figure}

The $Z$+jets background is modeled from an orthogonal control sample of events with a single photon produced in association with jets ($\gamma$+jets). This choice has the advantage of making use of a large statistics sample, which resembles the $\it {Z}$ production in all important aspects, i.e. production mechanism, underlying event conditions, pileup scenario, and hadronic recoil. Each $\gamma$+jets event should have at least one jet with $E_{T} >$15 GeV to reduce contamination from processes that have a photon produced in association with real $E_{T}^{miss}$, such as $W(l\nu) + \gamma$, $W(l\nu) +$ jets, where the jet is mismeasured as photon, and $Z(\nu\nu) + \gamma$ events. For matching the kinematics and overall normalization, an event by event reweighting as a function of dilepton $p_{T}$ has been applied in each jet bin separately. This takes into account the dependence of $E_{T}^{miss}$ on the associated hadronic activity. Residual discrepancies can arise due to the differences in the effective pile-up of the $\gamma$+jets sample due to the photon trigger pre-scale and event selection. These are taken into account by reweighting events according to the number of reconstructed vertices. This procedure produces an accurate model of the $E_{T}^{miss}$ distribution in $Z$+jets as can be seen in Fig.\ref{fig:MET_plots} which compares the $E_{T}^{miss}$ distribution of the reweighted $\gamma$+jets along with other backgrounds to the $E_{T}^{miss}$ distribution of the dilepton events in the data. To compute the $M_{T}$ for each $\gamma$+jets event, the value of $\vec{p_{T}}(ll)$ is defind as the photon $\vec{p_{T}}$ and the value of $M(ll)$ is chosen according to a probability density function constructed from the measured dilepton mass distribution in $Z$+jets events.

\subsection{{\bf Non-Resonant background Estimation}}
The background processes that do not involve a $Z$ resonance (non-resonant background) are estimated by using a control sample of events with two leptons of different flavor ($e^{+}\mu^{-}$/$e^{-}\mu^{+}$) that pass the full analysis selection. This background consists mainly of leptonic $\it {W}$ boson decays in $t\bar{t}$, $\it {tW}$ decays and $\it {WW}$ events. Small contributions from single top-quark events produced from s-channel and t-channel processes, $\it {W}$+jets events in which $\it {W}$ decays leptonically and a jet is mismeasured as lepton, and $Z\rightarrow \tau\tau$ events in which $\tau$ leptons produce light leptons and $E_{T}^{miss}$ are included in this estimate of the non-resonant background. This method cannot distinguish between the non-resonant background and a possible contribution from $H\rightarrow WW\rightarrow 2l2\nu$ events, which are treated as part of the non-resonant background estimate.\\
The non-resonant background in the $e^{+}e^{-}$/$\mu^{-}\mu^{+}$ final states is estimated by applying a scale factor ($\alpha$) to opposite flavor events: \\
\begin{equation}
N_{\mu\mu} = \alpha_{\mu} \times N_{e\mu}, N_{ee} = \alpha_{e} \times N_{e\mu} 
\end{equation}
Where $N_{ee}, N_{\mu\mu}$ and $N_{e\mu}$ are events passing the full selection in the $e^{+}e^{-}$, $\mu^{+}\mu^{-}$ and $e^{+}\mu^{-}$/$e^{-}\mu^{+}$ final states respectively.
The scale factor $\alpha$ is computed from the sidebands (SB) to the $Z$ peak (40 GeV/$c^{2}$ $<$$m_{H}$ $<$ 70 GeV/$c^{2}$ and 110 GeV/$c^{2}$ $<$ $m_{H}$ $<$ 200 GeV/$c^{2}$) using the following relations:
\begin{equation}
\alpha_{\mu} = \frac{N^{SB}_{\mu\mu}}{N^{SB}_{e\mu}}, \alpha_{e} = \frac{N^{SB}_{ee}}{N^{SB}_{e\mu}}
\end{equation}
where $N^{SB}_{ee}, N^{SB}_{\mu\mu}$ and $N^{SB}_{e\mu}$ are events in the sidebands in the $e^{+}e^{-}$, $\mu^{+}\mu^{-}$ and $e^{+}\mu^{-}$/$e^{-}\mu^{+}$ final states respectively. Such samples are selected by requiring $E_{T}^{miss} >$70 GeV and a $b$-tagged jet in the events. There is no requirement on jet multiplicity or VBF selection. The measured value of $\alpha$ with the corresponding statistical uncertainties are $\alpha_{\mu} = 0.61\pm0.04$ and $\alpha_{e} = 0.34\pm0.03$.

\section{Systematics and Results}
\begin{figure}[ht]
\begin{center}
\subfigure[]{\label{fig:Limit_plots-a}\includegraphics[angle=0,totalheight=.15\textheight,width=.30\textwidth]{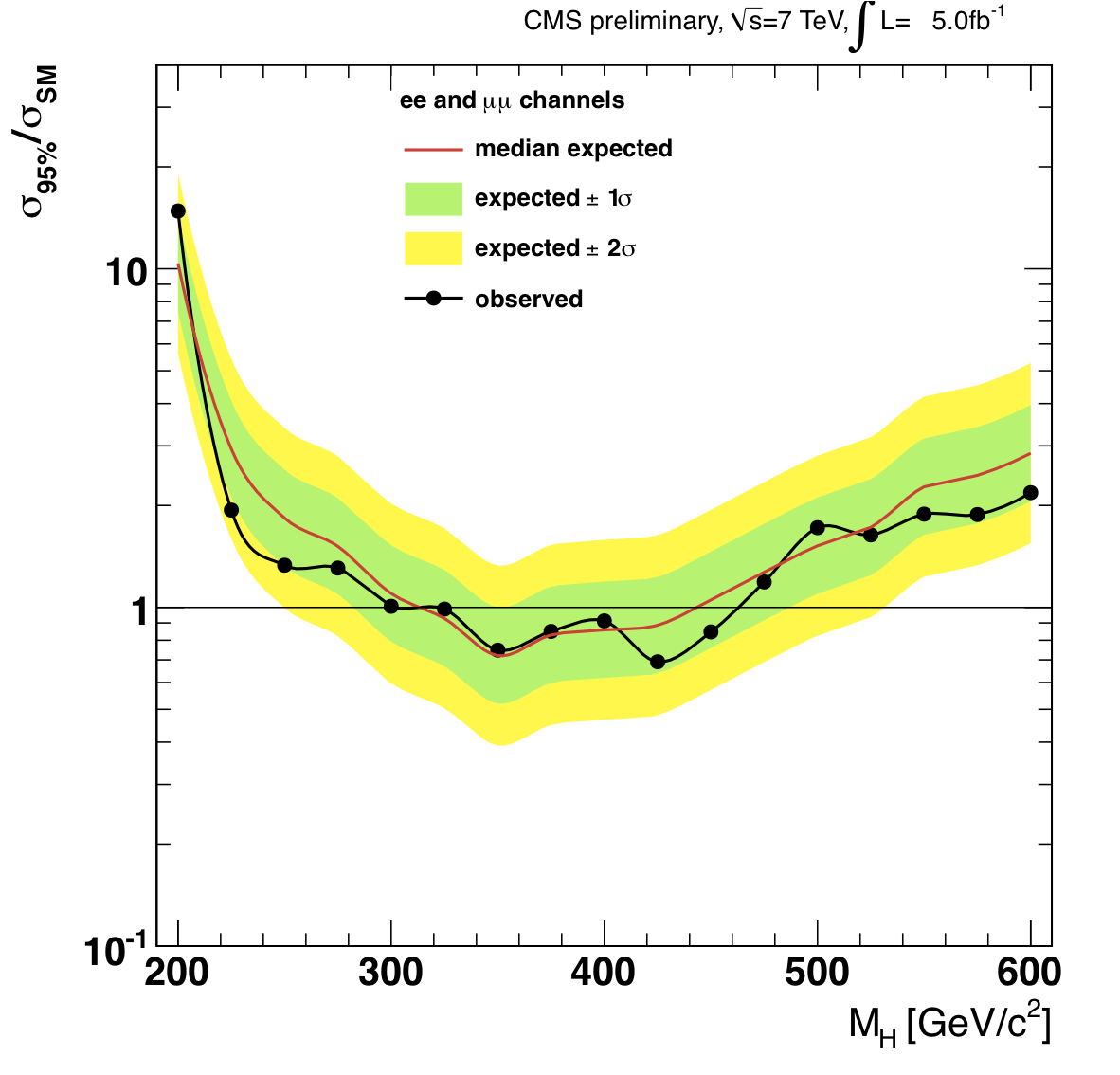}}
\subfigure[]{\label{fig:Limit_plots-b}\includegraphics[angle=0,totalheight=.15\textheight,width=.30\textwidth]{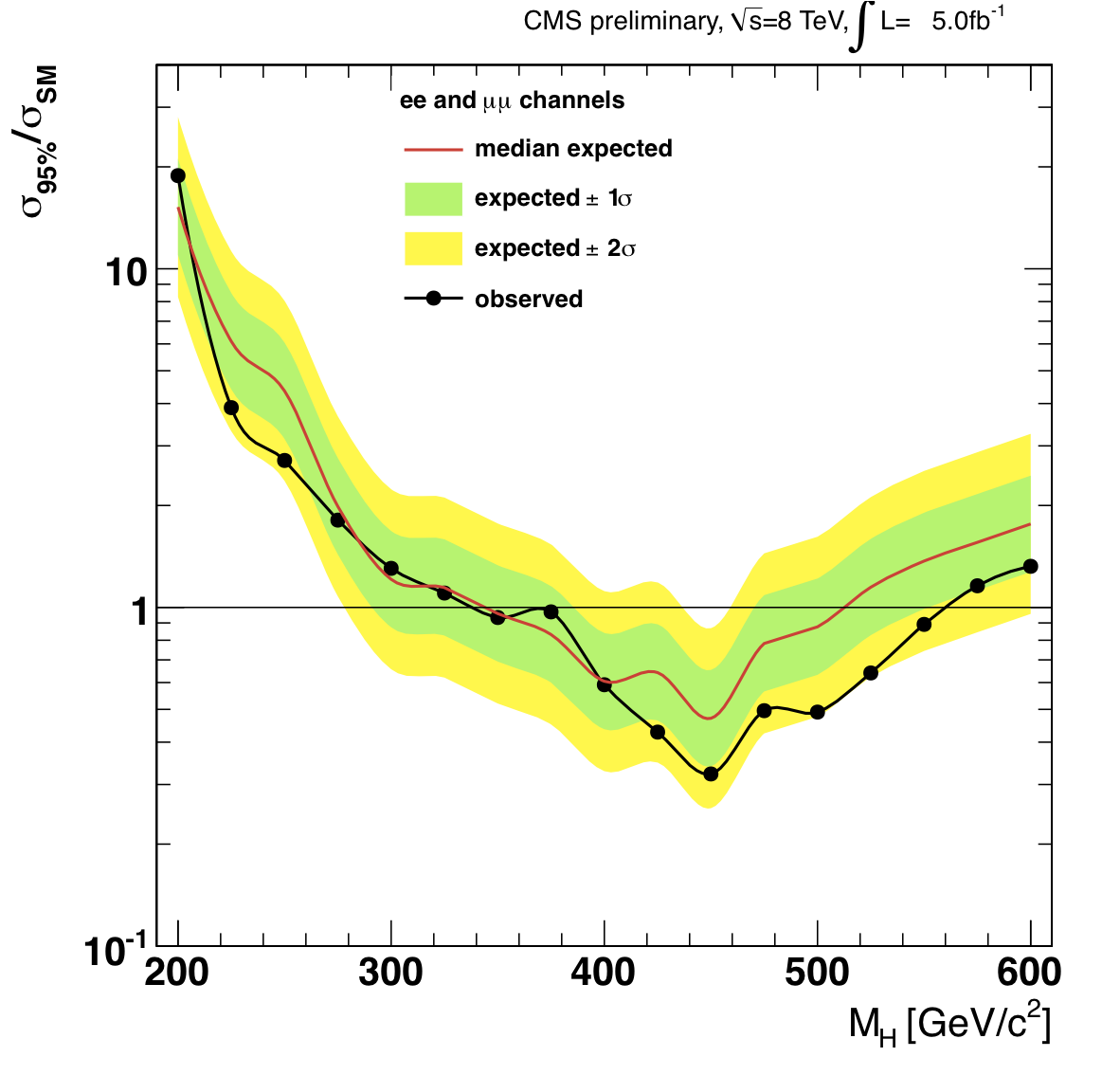}}
\subfigure[]{\label{fig:Limit_plots-c}\includegraphics[angle=0,totalheight=.15\textheight,width=.30\textwidth]{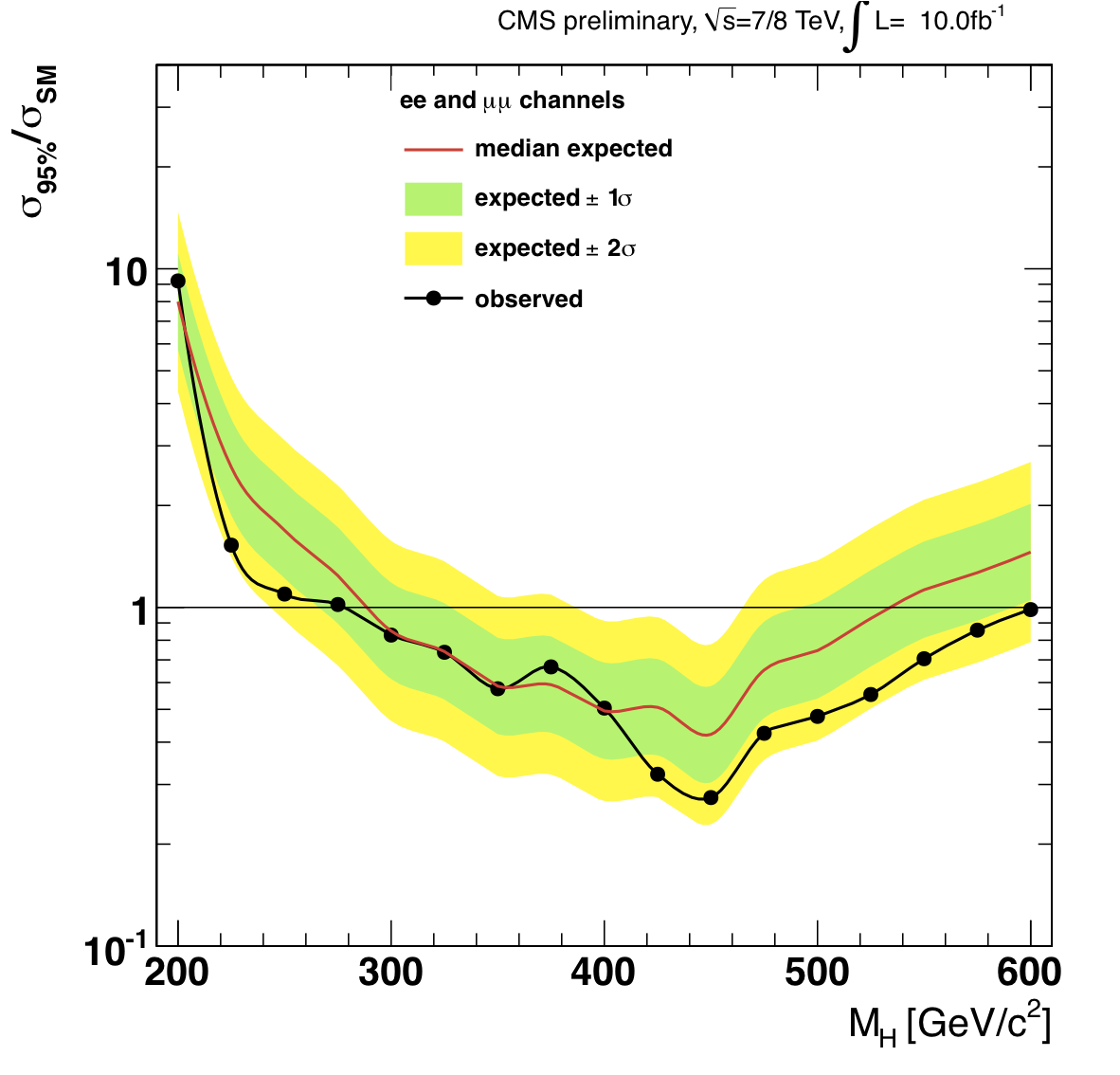}} \\
\caption{The ratio R of the 95\% CL upper limit cross section $\sigma$ to the SM Higgs boson production cross section $\sigma_{SM}$ as a function of $m_{H}$ for $\surd s$ = 7 TeV(a), $\surd s$ = 8 TeV(b) and for all data combined(c).}
\label{fig:Limit_plots}
\end{center}
\end{figure}

Systematic uncertainties include experimental uncertainties on the selection and measurement of the reconstructed objects, theoretical uncertainties on the signal and background processes which are derived from Monte Carlo simulation, and uncertainties on backgrounds determined from control samples in data. More details can be found in \cite{CMS-PAS-HIGS}.\\
The measured ratio R of the 95\% confidence level (CL) upper limit cross section $\sigma$ to the SM Higgs boson production cross section $\sigma_{SM}$ as a function of $M_{H}$ is shown in Fig. \ref{fig:Limit_plots} for $\surd s$ = 7 TeV and $\surd s$ = 8 TeV separately and also for both the data combined. The SM Higgs boson is excluded in the mass range 278-600 $GeV/c^{2}$ at 95\% CL, while the expected exclusion in the background-only hypothesis is 291-534 $GeV/c^{2}$.

\begin{thebibliography}{0}

\bibitem{CMS}
CMS Collaboration, ``The CMS experiment at CERN LHC'', {\it JINST} {\bf 3} (2008) S08004.

\bibitem{CMS-PAS-MUO}
CMS Collaboration,``Performance of muon identification in 2010 data'', {\it CMS Physics Analysis Summary} {\bf CMS-PAS-MUO-10-002 (2010)}.

\bibitem{CMS-PAS-EGM}
CMS Collaboration,``Electron reconstruction and identification at $\surd s$ = 7TeV'', {\it CMS Physics Analysis Summary} {\bf CMS-PAS-EGM-10-004(2010)}.

\bibitem{rho_cite}
M.Cacciari and G.P.Salam, ``Pileup subtraction using jet areas'', {\it Phys.Lett.B} {\bf 659}(2008)119, doi:10.1016/j.physletb.2007.09.077 arxiv:0.707.1378

\bibitem{CMS-PAS-JME}
CMS Collaboration, ``Jet Performance in pp Collisions at $\surd s$ = 7TeV'', {\it CMS Physics Analysis Summary} {\bf CMS-PAS-JME-10-003(2010).}

\bibitem{jhep_antikt}
M.Cacciari,G.P.Salam, and G.Soyez, ``The Anti-k(t) jet clustering algorithm'', {\it JHEP} {\bf 0808}(2008)063, doi:10.1088/1126-6708/2008/04/063, arxiv:0802.1189.

\bibitem{CMS-PAS-PFT}
CMS Collaboration, ``Particle-Flow Event Reconstruction in CMS and Performance for $E_{T}^{miss}$,Jets,Taus'', {\it CMS Physics Analysis Summary} {\bf CMS-PAS-PFT-09-001(2009).}

\bibitem{btag_cite}
CMS Collaboration, ``Measurement of the b-tagging efficiency using ttbar events'', {\it CMS Physics Analysis Summary} {\bf CMS-PAS-BTV-11-003(2011).}

\bibitem{CMS-PAS-HIGS}
CMS Collaboration, ``Search for the standard model Higgs boson in the $H\rightarrow ZZ\rightarrow 2l2\nu$ channel in pp collisions at $\surd s$ = 7TeV and $\surd s$ = 8TeV'', {\it CMS Physics Analysis Summary} {\bf CMS-PAS-HIG-12-023(2012)}

\end{thebibliography}
\end{document}